# Spectropolarimetry of the nova-like variable V1315 Aquilae

V. S. Dhillon[1] and R. G. M. Rutten[1,2]

[1] *Royal Greenwich Observatory, Apartado de Correos 321, Santa Cruz de La Palma, 38780 Tenerife, Canary Islands, Spain*
[2] *Netherlands Foundation for Research in Astronomy, P.O. Box 2, 7990 AA Dwingeloo, The Netherlands*

**ABSTRACT**
We present spectropolarimetric observations of the eclipsing nova-like variable V1315 Aql, obtained with the aim of determining whether the single-peaked uneclipsed lines observed in this and related nova-likes are the result of disc emission scattered into the line of sight by the wind. The data show linear polarization with a mean value of $0.11\pm0.02\%$. There are no significant differences between the continuum and line polarizations and no significant variations with wavelength or binary phase. We argue that the measured polarization may be attributed to scattering in the interstellar medium and hence conclude that there is no evidence of polarization intrinsic to V1315 Aql. We discuss alternative models which promise to resolve the controversy surrounding these objects.

## 1 INTRODUCTION

Cataclysmic variables (CVs) are interacting binary systems in which a white dwarf primary accretes material from a red dwarf secondary via an accretion disc (see the review by Wade & Ward 1985). V1315 Aquilae is an eclipsing member of the nova-like sub-class, whose members have never been observed to under-go nova or dwarf-nova type outbursts. Recent years have seen the emergence of a sub-group of nova-like variables which exhibit remarkably similar behaviour. To date, there exist five such well-studied systems: V1315 Aql (Downes et al. 1986; Dhillon, Marsh & Jones 1991), SW Sex (Penning et al. 1984; Honeycutt, Schlegel & Kaitchuck 1986), DW UMa (Shafter, Hessman & Zhang 1988; Dhillon, Jones & Marsh 1994), BH Lyn (Thorstensen, Davis & Ringwald 1991; Dhillon et al. 1992) and PX And (Thorstensen et al. 1991; Still, Dhillon & Jones 1995). All five objects are eclipsing systems with orbital periods lying in the narrow range of 3.24 to 3.74 hr; all five objects exhibit single-peaked Balmer and HeI emission lines which remain largely unobscured during primary eclipse and which show strong absorption features around the inferior conjunction of the emission-line source; all five objects show single-peaked high-excitation emission lines which are eclipsed by the secondary; and all five objects show radial velocity curves with significant phase shifts relative to photometric conjunction.

In addition, a number of systems are believed to be related to these five, exhibiting some, but not all, of the above features: WX Ari (Beuermann et al. 1992), UU Aqr (Diaz & Steiner 1991) and LB1800 (Buckley et al. 1990).

Thorstensen et al. (1991) coined the term 'SW Sex stars' to describe this group of objects. To date, no theoretical model has been able to unambiguously satisfy the observational constraints imposed by these systems (Dhillon, Marsh & Jones 1991). One possible solution to the problem, first proposed by Honeycutt, Schlegel & Kaitchuck (1986), invoked the existence of a wind region extending above and below the accretion disc, which remains visible during primary eclipse. However, although the abundant IUE observations of ultraviolet lines from the wind lends credence to a wind model, their high ionization implies that the wind can contribute little to the Balmer lines (Hoare 1994). An alternative solution, proposed by Rutten & Dhillon (1992), suggests that rather than being produced in the wind, the optical light is produced in the accretion disc and scattered into the line of sight by the wind. The scattered component would effectively mirror the accretion disc as viewed at a low-inclination. Hence this model would explain the single-peaked emission lines of SW Sex stars as the superposition of a double-peaked high-inclination accretion disc line (viewed directly from the disc) and a single-peaked low-inclination accretion disc line (viewed as scattered light from the wind).

This model would also explain the uneclipsed single-peaked lines, since during primary eclipse the continuum and the double-peaked lines from the disc are obscured, leaving only the single-peaked lines and suppressed continuum scattered by the uneclipsed wind. The other features characteristic of SW Sex stars, such as distorted radial velocity curves and phase 0.5 absorption might also plausibly be explained within such a model as due to the velocity field within the wind and self-absorption effects, respectively.

One way of testing this hypothesis is by observing an SW Sex star in polarized light (Rutten & Dhillon 1992). The bi-polar wind material will be irradiated by the disc from each side, creating a strongly asymmetric geometry. Viewed at a high inclination, the scattered light component will be strongly polarized and the polarized flux will reflect the spectrum of the irradiating source, ie. the accretion disc as viewed at a low inclination. If the wind is optically thin, opacities at the wavelengths corresponding to strong lines are much larger than in the continuum, thus giving rise to relatively strong scattering of line photons as opposed to continuum photons. Hence a polarization spectrum with strong single-peaked lines and a suppressed continuum would emerge.

In this paper we present the results of an attempt to test the above model by obtaining spectropolarimetry of the SW Sex-type star, V1315 Aql.

## 2   OBSERVATIONS AND DATA REDUCTION

V1315 Aql was observed on the night of 1993 July 26/27 with the 4.2 m William Herschel Telescope on La Palma. The blue arm of the ISIS triple-spectrograph (see Carter et al. 1993) was used, equipped with a thinned Tektronix CCD detector. A 158 lines mm$^{-1}$ grating gave a dispersion of 2.88 Å pixel$^{-1}$ and a wavelength range of $\lambda\lambda 3800 - 6700$ Å. The ISIS spectrograph is fitted with polarization optics for conducting linear spectropolarimetry (see Tinbergen & Rutten 1992). The optics consist of a half-wave plate located above the slit and a calcite plate analyser immediately below the slit. The calcite plate produces two slit images; the ordinary (o) and the extra-ordinary (e) ray, which are slightly offset in the spatial direction. A comb-type dekker mask with 4 arcsec apertures was used to allow simultaneous detection of object and sky without confusion between the o and e rays from different parts of the slit.

One component of the polarization vector of the incoming beam (eg. the Stokes Q parameter in some instrumental reference frame) is converted to a contrast in the intensities of the o and e ray spectra. Since both the o and e spectra are taken under exactly the same conditions, the ratio of the normally-extracted spectra is independent of sky transparency, seeing, image wander and variations in shutter timing. In order to account for differences in the response of the spectrograph and the detector to the polarized o and e rays, a second exposure with the half-wave plate rotated by 45° is required. This offset results in a rotation of the incoming polarization vector by 90°, which inverts the contrast between the o and e ray; the spectrograph response is then removed by taking the ratio of these two exposures. To measure the full polarization vector (ie. both Stokes Q and U) a second set of exposures is required with the half-wave plate set at 22.5° and 67.5°. Sky is removed by subtracting the mean spectrum of the sky regions on either side of the object spectra. This, to first order, also eliminates the influence of scattered light from the object (see Tinbergen & Rutten 1992 for details). Comparison arc spectra were taken between each set of polarization measurements in order to calibrate the wavelength scale and instrumental flexure.

The next step in the data reduction is to remove the instrumental zero-point from the polarization spectrum. This was achieved by observing the unpolarized calibration star HD154892 (Turnshek et al. 1990). We found no evidence for instrumental polarization at a level greater than 0.01% and therefore did not apply any correction. Similarly, by observing a calibration lamp through a Polaroid polarization filter we determined the efficiency of the system to be greater than 99% and therefore applied no correction to the degree-of-polarization scale. As an independent verification of the above, we observed the polarized standard star HD154445 (Hsu & Breger 1982) and found its measured polarization and wavelength dependence were in excellent agreement with its tabulated values. We are therefore confident of the quality of the observational and reduction techniques. A journal of observations is given in Table 1 for each set of four exposures yielding one polarization measurement.

## 3   RESULTS

The upper panel of Figure 1 displays the summed spectrum of V1315 Aql. Note that the eclipse has been included in the sum and the orbital motion has not been removed. No corrections for instrumental response and atmospheric losses have been made and therefore the shape and level of the continuum remains uncalibrated. The summed spectrum appears very similar to the ones displayed by Downes et al. (1986) and Dhillon, Marsh & Jones (1991), exhibiting single-peaked Balmer, He I, He II, C II and C III/N III emission lines. In addition, there are absorption lines of Ca II K, Fe II (or possibly Mg I) and Na I D clearly visible in the summed spectrum. Such features have also been observed in the nova-like PX And (Thorstensen et al. 1991), who measured their equivalent width and radial velocity variations. They found maximum absorption occured around phase 0.5 and the radial velocity curve was approximately in phase with the emission lines, but with different systemic and semi-amplitude velocities. A similar result was obtained by Szkody & Piché (1990) in their analysis of the phase 0.5 absorption feature in the Balmer lines and by Smith et al. (1993) in their analysis of the O I absorption feature near 7773Å. Our observations of these absorption features, although of much lower spectral and phase resolution, are in broad agreement with the above results.

The lower panel of Figure 1 displays the mean linear polarization spectrum of V1315 Aql, calculated from the same dataset as was used to produce the summed spectrum displayed in the upper panel. Two curves are presented in Figure 1 and care should be taken in their interpretation. The solid curve has been binned to yield errors of 0.10% in each interval. This curve indicates that there are no significant differences between the line and continuum polarizations and there are no gross variations with wavelength. However, since the noise associated with the normalized Stokes parameters always enhances the polarization value, the mean level is biased towards a higher value (Clarke & Stewart 1986). To determine the mean level we therefore binned the data to yield a single bin spanning the whole wavelength range (with a polarization error of 0.02% – this is a statistical value and does not include systematic errors). The result is shown by the dotted curve in Figure 1, which gives a mean polarization for V1315 Aql in this wavelength range of 0.11%. The polarization values determined from the four individual sets of observations over the orbital period are equivalent to within the measurement errors, and hence there is no evidence for any dependence on binary phase. It should be noted, however, that our observations do not completely rule out the possibility of variable polarization from V1315 Aql, since the observational technique we employed to measure linear spectropolarimetry smears out the polarization signal due to the long time interval required to make a measurement (approximately 40 mins; see Table 1).

## 4 DISCUSSION

The results of our observations show that there is evidence for only very low levels of polarization in the spectrum of V1315 Aql. This polarization could be either intrinsic to the system or due to interstellar absorption by dust grains aligned in the galactic magnetic field. To estimate the fraction of the polarization due to interstellar material we can use the result of the polarization survey by Serkowski, Mathewson & Ford (1975), which provides an estimate of the maximum level of interstellar polarization as a function of colour excess: $P_{max} \leq 9.0 E(B-V)$. The colour excess of V1315 Aql can be estimated from IUE spectra of the interstellar absorption bands around 2200Å, which gives a value of $E(B-V) = 0.0$ (Rutten, van Paradijs & Tinbergen 1992). An alternative estimate can be provided by calculating $E(B-V)$ from the distance of V1315 Aql (300 pc; Rutten, van Paradijs & Tinbergen 1992), which gives a (less accurate) mean value of $E(B-V) = 0.14$ (Scheffler 1967). Adopting $E(B-V) = 0.07$, this yields $P_{max} = 0.63\%$, which is sufficiently high to explain some, if not all, of the observed polarization. The weak dependence of interstellar polarization on wavelength (Serkowski, Mathewson & Ford 1975) is in general agreement with the observed data (see Figure 1). In conjunction with the fact that there is no evidence of orbital variability in the polarization signal, the above arguments lead us to conclude that the low level of linear polarization we have measured in V1315 Aql is consistent with the expected behaviour of the insterstellar medium and hence that there is no evidence for any polarized emission intrinsic to V1315 Aql.

This result does not necessarily discount electron scattering of disc emission in accretion disc winds as the cause of the uneclipsed, single-peaked lines observed in V1315 Aql and other nova-likes. It is possible that a number of different mechanisms act to weaken the measured signal. Firstly, multiple scatterings within the wind may destroy the polarization since the orientations of the polarization vectors are randomized and the geometry of the original radiation field is lost. Secondly, the wind might be spherical rather than bipolar (although ultraviolet observations argue against this;

Drew 1987), in which case the irradiation of the scattering particles occurs equally from all sides and the collective result of all the scattering processes results in zero polarization. Such a geometry would also affect the spectral shape, making the spectrum more blue in the case of Rayleigh scattering. Thirdly, collisions within the gas of scattering particles might reduce the polarization due to disturbance of the atom. A fine example of this is presented by Stenflo, Baur & Elmore (1980) in which of two adjacent spectral lines in the solar limb spectrum, one line is strongly polarized while the other line, which is formed where the collision rate is high, remains unpolarized. Finally, any polarization resulting from scattering might be diluted by radiation produced in an accretion disc wind; recent modelling has suggested that significant Balmer emission might be produced in a wind with a non-radial outflow (Hoare 1994) or from high-density clumps embedded in the wind (Marsh & Horne 1990).

Unfortunately, we do not appear to be any closer to identifying a model which can explain the various phenomena exhibited by V1315 Aql and related nova-likes (Dhillon, Marsh & Jones 1991). Given the numerous observations of these systems which now exist in the literature, it seems likely that the most fruitful line of inquiry will be from theoretical modelling. In particular, further modelling of optical line emission from CV winds with rotation and bipolar non-radial geometries would be desirable (Hoare 1994). Another model, which has yielded encouraging initial results, is the bright-spot overflow model of Lubow (1989), as applied to the nova-like variable PX And by Hellier & Robinson (1994). Although unable to explain the eclipse behaviour of the lines, it seems likely that such a model may be responsible for at least part of the phenomena observed in these systems, and a more detailed model and its application to other systems is now required. Finally, modelling of the outer-disc rim, as recently performed by Horne et al. (1994), may help explain the phase 0.5 spectral absorption features and yield important information about the origin and nature of the absorbing material.

From the present study we may conclude that, whatever model is finally proposed to explain the SW Sex stars, it must not produce measureable linearly-polarized radiation.

ACKNOWLEDGEMENTS

We thank Dr. Rob Rutten for sharing his knowledge of photon scattering with us. The William Herschel Telescope is operated on the island of La Palma by the Royal Greenwich Observatory in the Spanish Observatorio del Roque de los Muchachos of the Instituto de Astrofísica de Canarias.

**Table 1.** Journal of observations. Each half-wave plate (HWP) cycle refers to one complete polarization measurement, consisting of four exposures at HWP angles of 0.0°, 45.0°, 22.5° and 67.5°. The UTC columns give the start and end times of a single HWP cycle of four exposures. The final two columns give the start and end orbital phases of each HWP cycle, calculated from the ephemeris of Dhillon, Marsh & Jones (1991).

| HWP cycle | Exposure time (s) | UTC start | UTC end | Phase start | Phase end |
|---|---|---|---|---|---|
| 1 | 4 x 600 | 22:11 | 22:55 | 23571.692 | 23571.909 |
| 2 | 4 x 600 | 22:56 | 23:40 | 23571.916 | 23572.133 |
| 3 | 4 x 600 | 23:53 | 00:36 | 23572.197 | 23572.415 |
| 4 | 4 x 600 | 00:39 | 01:23 | 23572.427 | 23572.644 |

**Figure 1.** Upper panel: Summed spectrum of V1315 Aql. Lower panel: Percentage linear polarization spectrum of V1315 Aql. The solid curve displays the polarization data binned to yield errors of 0.10% in each wavelength interval. The dotted curve represents the mean polarization level over the whole wavelength range.